\documentclass[preprint]{aastex}
\tighten 
\begin{document}
\newcommand{\lya}{Lyman~$\alpha$}
\newcommand{\lyb}{Lyman~$\beta$}
\newcommand{\za}{$z_{\rm abs}$}
\newcommand{\ze}{$z_{\rm em}$}
\newcommand{\cmtwo}{cm$^{-2}$}
\newcommand{\nhi}{$N$(H$^0$)}
\newcommand{\degpoint}{\mbox{$^\circ\mskip-7.0mu.\,$}}
\newcommand{\kms}{\,km~s$^{-1}$~}      % note leading thinspace
\newcommand{\minpoint}{\mbox{$'\mskip-4.7mu.\mskip0.8mu$}}
\newcommand{\peryr}{\mbox{$\>\rm yr^{-1}$}}
\newcommand{\secpoint}{\mbox{$''\mskip-7.6mu.\,$}}
\newcommand{\sqdeg}{\mbox{${\rm deg}^2$}}
\newcommand{\squig}{\sim\!\!}
\newcommand{\subsun}{\mbox{$_{\twelvesy\odot}$}}
\newcommand{\et}{{\rm et al.}~}

\def\ltsima{$\; \buildrel < \over \sim \;$}
\def\simlt{\lower.5ex\hbox{\ltsima}}
\def\gtsima{$\; \buildrel > \over \sim \;$}
\def\simgt{\lower.5ex\hbox{\gtsima}}
\def\arcs{$''~$}
\def\arcm{$'~$}
\def\erf{\mathop{\rm erf}}
\def\erfc{\mathop{\rm erfc}}
\title{THE KINEMATIC CONNECTION BETWEEN QSO-ABSORBING GAS AND GALAXIES AT INTERMEDIATE REDSHIFT
\altaffilmark{1,2}}
\author{\sc Charles C. Steidel\altaffilmark{3}, Juna A. Kollmeier\altaffilmark{4}, 
and Alice E. Shapley}
\affil{Palomar Observatory, California Institute of Technology, MS 105--24, Pasadena, CA 91125}
\author{\sc Christopher W. Churchill}
\affil{Pennsylvania State University, Department of Astronomy and Astrophysics, 525 Davey Lab, University
Park, PA 16802}
\author{\sc Mark Dickinson}
\affil{Space Telescope Science Institute, 3700 San Martin Drive, Baltimore, MD 21218}
\author{\sc Max Pettini}
\affil{Institute of Astronomy, Madingley Road, Cambridge CB3 OHA, UK}

\altaffiltext{1}{Based, in part, on data obtained at the 
W.M. Keck Observatory, which 
is operated as a scientific partnership among the California Institute of Technology, the
University of California, and NASA, and was made possible by the generous financial
support of the W.M. Keck Foundation.
} 
\altaffiltext{2}{Based, in part, on data obtained with the NASA/ESA Hubble Space
Telescope, which is operated by the Space Telescope Science Institute for the
Associated Universities for Research in Astronomy, Inc., under NASA contract
NAS5-26555.}
\altaffiltext{3}{Packard Fellow}
\altaffiltext{4}{Present address: Ohio State University, Department of Astronomy}
\begin{abstract}
We present complementary data on 5 intermediate redshift ($0.44 \le z \le 0.66$) Mg~II
absorbing galaxies, combining high spatial resolution imaging from {\it Hubble Space 
Telescope}, high--resolution QSO spectroscopy from Keck/HIRES, and galaxy kinematics
from intermediate resolution spectroscopy using Keck/LRIS. These data allow a direct
comparison of the kinematics of gas at large galactocentric impact parameters with
the galaxy kinematics obtained from the faint galaxy spectroscopy. 
All 5 galaxies appear to be relatively normal spirals, with measured
rotation curves yielding circular velocities in the range $100 \le v_c \le 260$ \kms. 
The QSO sightlines have projected impact parameters to the absorbing galaxies
in the range $14.5{\rm h}^{-1} \le d \le 75{\rm h}^{-1}$ kpc; the galaxies
have inclination angles with respect to the line of sight ranging from 
40 to 75 deg.
We find that in 4 of the 5 cases examined, the velocities of all of the Mg~II absorption components
lie entirely to one side of the galaxy systemic redshift. The fifth case, for which the galaxy
is much less luminous than the other 4, 
has narrow absorption centered 
at zero velocity with respect to systemic despite having the largest disk inclination
angle in the sample.  These observations are consistent with rotation being dominant  
for the absorbing gas kinematics; 
however, the total {\it range} of velocities observed is inconsistent
with simple disk rotation in every case. Simple kinematic models that simultaneously
explain both the systemic offset of the absorbing material relative to the galaxy
redshifts, {\it and} the total velocity width spanned by the absorption, 
require either extremely thick rotating gas layers, rotation velocities that
vary with $z$ height above the extrapolation of the galactic plane, or both. In any case,
our small sample suggests that rotating ``halo'' gas is a common feature of intermediate redshift
spiral galaxies, and that the kinematic signature of rotation dominates over radial
infall or outflow even for gas well away from the galactic plane. 
We discuss possible explanations for this behavior, and compare
our observations to possible local analogs. 
\end{abstract}
%\newpage

\section{INTRODUCTION}

Metallic absorption lines in the spectra of background QSOs and the nature of 
their association with galaxies have been subjects of considerable interest over
the last decade. 
A number of circumstantial pieces of evidence, accumulated
during the first $\sim 20$ years of QSO absorption line research, pointed
to galaxies as being responsible: e.g. the tendency for systems to split into
complexes of total velocity extent consistent with galaxy-sized potential wells,
the presence of metals which were presumably produced {\it in situ}, and clustering
properties that resembled that of galaxies. However, exploring the details
of the connection between absorption systems and galaxies did not become
possible until 
later work began to directly identify galaxies responsible for individual QSO
absorption line systems (Bergeron \& Boiss\'e 1991; Steidel, Dickinson,\& Persson 1994 (SDP);
Le Brun \et 1997).
This connection is extremely interesting
in the context of our understanding of galaxy evolution and galactic structure
because QSO absorption line systems have the potential to explore the gas-phase 
physical conditions, geometry, and kinematics of galaxies with what can be many
orders of magnitude increased sensitivity compared to  
direct observations of the galaxies. Moreover, the sensitivity of absorption line
measurements is largely independent of galaxy redshift, and thus provides the opportunity to
study the evolution, over very large time baselines, of characteristics
that can be much more subtle than those accessible using traditional faint galaxy
techniques. 

At present, most identification of absorption systems with individual galaxies is
tenuous at best. If a faint galaxy is found near the line of sight with a redshift
in agreement with
the measured redshift of the QSO absorption system, it is generally taken as
as positive identification of the actual object producing the absorption.  
This type of statistical approach has been adopted by most studies of intermediate
redshift metallic absorption systems (e.g., BB91; SDP; Le Brun \et 1997). 
There are very few cases where relatively exhaustive spectroscopy down to
faint magnitude limits has been obtained for all objects within several hundred
kpc of the QSO line of sight (cf. Steidel \et 1997), making the identifications
of each absorber more secure. Possible selection effects inherent in this type
of identification procedure have been outlined by Charlton \& Churchill (1996). 

The state of knowledge of the types of galaxies producing various classes of
QSO absorption systems varies considerably as a function of redshift and of
absorption system taxonomy. The most work has been done at $z \simlt 1$, for
absorption line systems selected by the presence of Mg~II $\lambda\lambda 2796$,
2803 doublets with absorption line  
rest-frame equivalent widths $W_{\lambda} > 0.3$ \AA\ (BB91; SDP, Steidel 1995). These
relatively strong Mg~II--selected systems are
generally associated with gas having N(H~I)$ \simgt 10^{17}$ cm$^{-2}$ (i.e., ``Lyman limit systems''),
and so would be expected to probe (on average) the outer parts of galaxies where the H~I
is more highly ionized than disk gas observed for nearby galaxies. 
The galaxies responsible for the Mg~II absorption at
intermediate redshift ($\langle z \rangle \simeq 0.6$), statistically,
appear to be drawn from normal field galaxies with luminosities within $\sim 1.5$ magnitude
of present--day $L^{\ast}$ (BB91; SDP94). More recent morphological studies using
{\it HST}, of which the data in this paper are a subset, have shown
that the identified galaxies are generally of relatively normal morphologies identifiable
along the Hubble sequence (Dickinson \& Steidel 1996; Steidel 1998, Steidel \et 1997). 
The colors and magnitudes
of these galaxies appear to exhibit little or no evolution with redshift over the range
$0.3 \simlt z \simlt 0.9$, in agreement with general studies of field galaxy
evolution (e.g., SDP, Lilly \et 1996; Vogt \et 1996).  There is
some evidence that the observed distribution of ``impact parameters'', the
projected physical distance between the putative absorbing galaxy and the QSO
sightline at the galaxy redshift, is consistent with a roughly spherical distribution
of gas of radius $R(L)$ that is a weak function of luminosity (SDP, Steidel 1995, 1998): 
$$R(L_K) \simeq 38h^{-1} kpc \left(L_K \over L^{\ast}_K\right)^{0.2}.$$ 
A flattened
disk--like geometry is not as favored by the existing statistics, but any
geometry in which the gas layer has finite thickness compared to its radial
extent is very difficult to rule out with present statistics on impact
parameters and the incidence of interlopers (cf. Charlton \& Churchill 1996). 
Given the available
statistics, however, and the likely stochastic nature of the sizes and
shapes of gaseous envelopes around field galaxies, simple geometric pictures should be 
viewed as no more than working models. 

Damped Lyman $\alpha$ systems, which are metal line systems for
which $N(HI) \ge 2 \times 10^{20}$ cm$^{-2}$ (cf. Wolfe \et 1986), have
also received a great deal of attention in the context of what their
kinematics can tell us about galaxies, particularly at high redshift.  
In a series of papers, Prochaska \& Wolfe (1997,1998) present analyses
of the kinematics of high redshift damped Lyman $\alpha$ systems ($z \simgt 2$),
concluding that the ``edge leading asymmetry'' often seen in the velocity 
profiles of DLAs is most consistent with rotating thick disks with $v_c \simgt
200$ \kms. Other work has claimed that the kinematics can be explained by
aggregates of small galactic fragments as would naturally be present in
hierarchical structure formation models (Haehnelt, Steinmetz, \& Rauch 1998) or
simply from randomly moving clouds in spherical halos 
(McDonald \& Miralda-Escud\'e 1999), although
all models appear to have problems reproducing the relative velocities
of the low--ionization and high ionization species in the damped Lyman $\alpha$
systems (Wolfe \& Prochaska 2000).  
At present, little or no information on the galaxies themselves
is available at these high redshifts, so that the disk inclination angles,
impact parameters, and galaxy systemic redshifts are generally unknown. 
At lower redshifts, in contrast to the
Mg~II absorbers, both ground--based and HST observations of the galaxies
indicate that the galaxies associated with the high column density absorbers 
are a very ``mixed bag'', ranging from dwarf galaxies as faint as
$0.05$ L$^{\ast}$ to normal spirals (Steidel \et 1994;
Steidel \et 1997; Le Brun \et 1997; Turnshek \et 2000; Bowen, Tripp, \& Jenkins 2001).  In many cases 
spectroscopy of the putative DLA absorbers has been difficult or impossible
because of the very small impact parameters (and resulting problems
with scattered light from the QSOs) that often obtain for DLAs. It should be emphasized
that every DLA is also a Mg~II system (cf. Rao \& Turnshek 2000), but that
the Mg~II systems are sensitive to gas with H~I column densities up to $\sim 3$ 
orders of magnitude smaller than that which will produce a DLA. 
It is certainly possible that disk--like kinematics could dominate
many Mg~II systems even when the impact parameter is well 
beyond the $10^{20}$ cm$^{-2}$ H~I contour. In any case, studies of systems where
the absorption line kinematics and the galaxy properties can be used simultaneously to constrain
kinematic models would be very valuable for understanding the DLA systems at
high redshift. For the moment, this can only be done at substantially lower redshifts ($z \simlt 1$).

The advent of echelle spectrographs on 8m--class telescopes has allowed
relatively routine high dispersion spectroscopy of the same QSOs that have
been used historically for Mg~II absorption surveys (e.g., Steidel \& Sargent 1992).
There has been a fair amount of activity in examining the kinematics of the 
intermediate redshift Mg~II systems (to scales as fine as $\sim 5$ \kms)
and comparing them with what is known
about the absorbing galaxy photometric properties and the observed
impact parameters (Lanzetta \& Bowen 1992; Churchill, Steidel, \& Vogt 1996; Churchill \& Vogt 2001). 
The relatively small samples
of absorption system/absorbing galaxy pairs studied so far do not provide any
obvious clues to the nature of absorbing gas in relation to the galaxies, or at
least no strong systematic trends. 
So far, the issues of the geometry, physical conditions,
and kinematics of the absorbing gas have not been considered in concert with simultaneous access
to high quality imaging and spectroscopic data on the faint galaxy absorbing candidates.  
And yet, it is now possible with 8m-class telescopes to obtain spectra of quality sufficient
to trace the kinematics of galaxy rotation curves to $z \sim 1$ (e.g., Vogt \et 1996), 
and it is clearly straightforward to obtain high quality images with kpc-scale resolution for
galaxies in the same redshift range using {\it Hubble Space Telescope} (HST). 

In this paper, we present pilot observations that provide first
results and explore the feasibility of establishing directly the kinematic connection
between gas at large galactocentric impact parameters, and the luminous component of
the galaxies.  At the very least, it should be possible to test the hypothesis that the absorbing
gas is dynamically associated with the identified absorbing galaxy using greatly
improved redshift and kinematic measurements of both the absorbing gas and the luminous
material in the galaxy. At best, we hope to establish the nature of the absorbing gas; some
possibilities include:

\begin{itemize}
\item The gas is an extension of the galaxy disk, and the kinematics of the absorbing
material are entirely compatible with a kinematic extension of the observed galaxy
rotation curve. If this is the case, there may be an opportunity for measuring 
galaxy rotation curves out to much larger galactocentric radii than is normally
possible, particularly for galaxies at relatively high redshift.

\item The velocities of the galaxy and the absorption are too discrepant for
the gas to be plausibly associated with the identified galaxy; the true absorber
may remain unidentified.

\item The galaxy and gas-phase kinematics exhibit the same general kinematic
spread and zero net offset, as might be expected if the absorbing gas 
takes the form of distributed ``clouds'' or substructure with random velocities
relative to systemic

\item Some combination of the above.
\end{itemize}

The paper is organized as follows. In \S 2 we present the data. \S 3 contains a general
discussion of results in each of the 5 fields, and \S 4 explores simple kinematic models
to explain the observations. Finally, in \S 5 we discuss the general results and their
implications.

\section{DATA}

The QSO/absorbing galaxy fields observed in this pilot program were chosen from among
the 25 fields of intermediate redshift Mg~II--selected absorption line systems
we observed with {\it Hubble Space Telescope}. All of these QSO fields have been included
in existing analyses of the statistics of Mg~II absorbing galaxies (e.g., Steidel, Dickinson, \& Persson 1994)
and most were observed as part of a large low-resolution survey for $0.2 \le z \le 2.2$ Mg~II
absorption systems (Steidel \& Sargent 1992). More details are given in \S 3.  

The initial targets for this project were {\it not} chosen at random from among the Mg~II
absorbing galaxies at redshifts where [OII] $\lambda 3727$ would be accessible at
optical wavelengths. Rather, they were chosen to have apparent brightnesses and morphologies
that would allow for interesting galaxy kinematics. Most of them have large inclination
angles with respect to the line of sight, and all but 1 are brighter than ${\cal R}$=22,
so that very high quality spectra could be obtained without large investments in observing time. 
Choosing galaxies with relatively large inclination angles also (in principle) allows
extending kinematic measurements of the galaxy rotation curves well beyond the luminous
portions using the absorption line kinematics, which should be sensitive to H~I column
densities perhaps 100 times smaller than most 21-cm measurements of galaxies in the local
universe. 

While it is generally the case that the galaxies identified as Mg~II absorbers
at intermediate redshifts are classifiable along the full range of the Hubble sequence (Steidel 1998), 
in the case of the 5 galaxies in this pilot sample, 
all appear morphologically to be normal mid-to-late--type spiral galaxies,
albeit at redshifts of $\sim 0.5$. All but one are close to present--day $L^{\ast}$ in rest-frame
blue luminosity (see Table 1). 

\subsection{Ground-based Imaging}

Optical and near--IR photometry of the galaxies of interest was available from ground-based
observations obtained primarily during the years 1991-94 as part of a more extensive survey
of Mg~II absorbing galaxies (Steidel, Dickinson, \& Persson 1994).  
The data for the 5 fields discussed in this paper were obtained using the Kitt Peak National
Observatory 2.1m telescope (optical) and 4m Mayall telescope (near-IR).   
Some of the relevant data are collected in Table 1. The optical ${\cal R}$ passband,
through which all of the fields were observed, has an effective wavelength of 6930 \AA\ and
thus is a very close approximation to rest--frame B at a redshift of $z=0.5-0.6$. The
${\cal R}$ magnitudes were converted to rest--frame B absolute magnitudes using k-corrections
appropriate to the best--fit spectral type from the optical and optical/IR colors. These k-corrections
were in all cases smaller than 0.2 mag.  With the exception of G1 1222+228, which has a luminosity of
only $\sim 0.25$L$^{\ast}$, all of the galaxies are comparable to or more luminous than the present--day
$L^{\ast}$ of $M_B \simeq -19.5$ (Folkes \et 1999; Blanton \et 2001).  

Spectroscopic confirmation of the absorbing galaxies in this sample had been obtained
previously for G1 1038+064 (Bergeron \& Boiss\'e 1991), and for the other galaxies
through our own spectroscopic efforts using the Lick Observatory Shane 3m telescope
in the 1991-93 observing seasons. The spectroscopic identifications of G1 1222+228 and
G1 1317+276 were both tentative due to poor-quality spectra; as discussed in \S 3 below,
the Keck spectroscopy led to substantial revisions of the absorbing galaxy situation
in those two cases. 

Also listed in Table 1 are the angular separation of the galaxies from the QSO, in arc seconds, and
the projected impact parameter in $h^{-1}$ kpc. 

\subsection{HST Imaging}

Images of each of the QSO fields discussed in this paper were obtained using WFPC-2 during
Cycles 5 and 6 as part of a more general HST imaging survey of QSO absorbing galaxies
at intermediate redshifts. The details of the  observations are summarized in Table 2. 
In brief, each field was observed in the F702W filter (again, in order to be well-matched to
rest-frame B at the typical absorber redshift of $z_{abs}\simeq 0.6$, and for compatibility
with our existing ground-based data discussed above) for two orbits, with
standard CR-split exposures during each orbit and a non-integral pixel dither
between orbits to allow for partial recovery of the spatial resolution that is degraded
by under-sampling. The images were reduced using the ``variable pixel linear combination''
(or ``drizzling'') technique (Fruchter \& Hook 1997) onto a final pixel scale of
0\secpoint05 pixel$^{-1}$. In each case, the QSO was centered on the Wide Field Camera
chip 3; portions of the reduced images, with the galaxies of interest marked, 
are shown in Figures 1-5. Galaxy inclination angles $\theta_i$ with respect to the plane
of the sky, resulting from fits to the galaxy isophotes, are given in Table 2.

\subsection{Galaxy Spectroscopy}

Spectra of the absorbing galaxies were obtained in 1999 March using LRIS (Oke \et 1995) on the
Keck II telescope. As detailed in Table 3, 3 of the galaxy spectra were obtained with a 1\secpoint0
long slit. The position angle of the slit was chosen to lie along the major axis of the
galaxy of interest (in the case of G1 Q1222+228), to include 2 galaxies near the QSO sightline
in the case where there was ambiguity about the identification of the absorber (as for G1,G2 Q1317+2743),
or in order to minimize to leakage of scattered QSO light into the slit (as for Q1148+3842). 
For the two other cases (G1 Q0827+2421 and G1 Q1038+0625) the observations were obtained through
a slit mask (also having 1.0 arc sec slitlets) whose position angle was chosen so that
the slit of primary interest lay along the major axis of the absorbing galaxy. A 600 line/mm
grating blazed at 7500 \AA\ and tilted so as to place redshifted [OII] $\lambda 3727$ emission
near the center of the spectral format, was used for all of the observations. With the 1.0 arc sec
slits, the resulting spectral resolution was 4.5 \AA\, or $\sim 225$ \kms at the wavelengths
of interest for measuring [OII] $\lambda$ 3727 at $z \simeq 0.6$. 

The data were flat-fielded using exposures of a halogen lamp internal to the spectrograph. 
Sky subtraction was accomplished by fitting a polynomial function to each spatial column
of the spectrogram, in the usual fashion. Extended line emission was then evident for each
galaxy over a spatial region of $\sim 2$ arc seconds along the slit. We extracted
individual one-dimensional spectra by co-adding 3-pixel swaths (which corresponds
to approximately one resolution element of $\simeq 0.65$ arc seconds) at one pixel (0.215 arc seconds) 
increments along the slit (so that adjacent spectra are not completely independent of one another). 
Wavelength calibration, which is crucial given our interest in an external comparison of measured velocities, 
was accomplished by extracting spectra of HgArNeKr arc line lamps at the same spatial pixels
as the extracted galaxy spectra. 
In order to account for the possibility of instrumental flexure between the time the data
and the arc lamps were obtained, we adjusted the wavelength solutions for each individual galaxy exposure (generally there were
two exposures, each of 1200s duration, with a spatial dither along the slit in between) using
night sky emission lines. The RMS residuals of the wavelength solutions were $\sim 0.2$ \AA, or
about 10 \kms at the wavelengths of interest. Wavelengths were reduced to heliocentric to facilitate
the comparison with the absorption line kinematics. 

Rotation curves were measured for each galaxy by fitting the position of the [OII] doublet
at each spatial position over the extent of the galaxy, in a manner similar to that
described in Vogt \et 1996. Uncertainties in the emission line velocities were estimated
from the RMS uncertainties produced by the fitting procedure; these uncertainties are
typically 25 \kms per spatial point, or about one tenth of a spectral resolution element. 
The systemic redshift of each galaxy was taken to be the velocity at the position
of the centroid of the continuum (averaged over one spatial resolution element) on the 2-dimensional spectrogram. 
The resulting rotation curves are plotted for each galaxy in Figures 1-5. 

Circular velocities $v_c$ were estimated for each galaxy by assuming that the extreme observed velocity
with respect to systemic is equal to $\rm v_c sin \theta_i$. These estimates must be viewed as
approximate, since it is not obvious that the observed rotation curves extend far enough to
have reached their asymptotic values, and because of uncertainties in $\theta_i$. In any case,
our conclusions are not strongly affected by lack of precision in the estimates of $\rm v_c$.

\subsection{QSO Spectroscopy}

Spectra of the 5 QSOs were obtained using the HIRES echelle spectrometer on the Keck I telescope (Vogt \et 1994),
all using a 0\secpoint85 slit resulting in spectral resolution of 6.6 \kms. Three of the QSOs
were observed in 1995 January, with the details of
the observations and reductions described in Churchill \& Vogt (2001). Two of the QSOs were 
observed in 1998 February and March, one using the blue--blazed cross-disperser (Q0827+2421). 
The echelle and cross-disperser angles were chosen so as to include at least the Mg~II $\lambda\lambda 2796$,2803 doublet
at the redshift of a known QSO absorbing galaxy; in most cases several other low-ionization transitions
were also observed. 

The 1998 HIRES data were reduced using MAKEE (T. Barlow, private communication), a package tailored
to the reduction of HIRES data. The output of MAKEE is an extracted spectrum of each echelle order, corrected
for the echelle blaze function, and transformed to vacuum, heliocentric velocities.  
A summary of the HIRES observations is provided in Table 4.  Relevant portions of the HIRES
spectra are shown in Figures 1-5.

\section{Discussion of Individual Fields}

\subsection{Q0827+243 (OJ 248)}

The absorption system at $z_{abs}=0.52499$ is among the strongest Mg~II systems known, with
the $\lambda 2796$ component having a rest--frame equivalent width of 2.47 \AA. The Mg~II 
absorption is strongly saturated, with a total velocity width of $\simeq 270$ \kms; some
insight into the kinematic structure is possible by looking at the apparently unsaturated
Mg~I $\lambda 2853$ line, which shows at least 4 components of roughly equal
strength spread over the entire velocity range covered by the Mg~II ``trough''. 
This system is known to be a DLA (Rao \& Turnshek 2000), with a measured
H~I column density of $2 \times 10^{20}$ cm$^{-2}$ (which could include gas over
the full $\sim 300$ \kms velocity range).
 
The spectrum of the absorbing galaxy, which was obtained with the slit oriented
along the major axis of the galaxy, yields a rotation curve over the velocity range
$-180 \le v_{gal} \le 240$ \kms, but the blue-shifted side of the galaxy is clearly
affected by a satellite galaxy that appears to be distorting the galaxy disk and
which has a systemic velocity different from the galaxy of interest (skewed toward
positive velocities, as can be seen in the top panel of Figure 1b). It is unclear
how or if this apparent satellite is affecting the kinematics of the absorbing
gas---there is no Mg~II absorption at positive velocities, but it is possible
that some of the kinematic complexities of the gas may be induced by an imminent
merger event. Because of this distortion to the galaxy morphology (whatever
its cause), the inclination
angle of the galaxy (measured to be $i=69^{\circ}$) is rather uncertain.  
Assuming this inclination angle, the galaxy rotation speed is $v_c \simeq 260$ \kms,
evaluated from the side of the rotation curve that is unaffected by the satellite
(i.e., at positive velocities with respect to systemic).
Unlike the other 4 galaxies discussed in this paper, G1 0827+243 has very strong
[OII] emission ($W_{\lambda}^0\simeq 50$ \AA) 
characteristic of vigorous current star formation. The rest-frame
B luminosity of $\sim 1.6$ times present--day $L^{\ast}$, may be significantly
enhanced by this star formation. 

The absorbing gas kinematics are qualitatively as would be expected for 
a model in which the extrema of the absorbing gas velocities are consistent
with a kinematic extension of the disk gas seen
in emission, but extending to somewhat larger velocities at an impact parameter
of $25.4$h$^{-1}$ kpc. We discuss more detailed kinematic models for this
system in \S 4.1.    

\subsection{Q1038+064 (4C 06.41)}

The absorption system at $z_{abs}=0.4415$ has been known for more than 20 years
(Burbidge \et 1977; Weymann \et 1979) and the absorbing galaxy was among the
first intermediate-redshift systems identified (Bergeron \& Boisse\'\ 1991).
The {\it HST} image in Figure 2a clearly shows that the absorbing galaxy is a
luminous but relatively normal mid-type spiral, with an inclination angle of $i=60^{\circ}$. 
The rotation curve of the galaxy is well--determined, with an observed rotational
velocity of $\rm v_{c} sin \theta_i \simeq 225$ \kms, or a de-projected rotation speed
of $v_{c}^{corr} \simeq 260$ \kms. The absorbing gas follows the kinematics of
the emitting material nicely, indicating an extreme velocity of $\simeq 250$ \kms
relative to the galaxy systemic velocity, with the sign as expected for a simple
extension of the rotation curve to a disk impact parameter of $44.8$h$^{-1}$ kpc. 
If the absorbing gas is interpreted as an extension of a flat rotation curve to
a galactocentric radius of $\simeq 45$h$^{-1}$ kpc, the minimum implied virial
mass of the galaxy is $\simeq 7{\rm h}^{-1} \times 10^{11}$ M$_{\sun}$. We discuss
the kinematic model for this system in \S 4.2. 

The Mg~I $\lambda 2853$ absorption is very weak, with only one component securely
detected at $v_{sys}=-120$ \kms and a marginal detection of the $v_{sys}\simeq -180$
\kms component that is the strongest in Mg~II. Inspection of an archival HST 
Faint Object Spectrograph spectrum of Q1038+064 reveals a Lyman limit system
at a redshift compatible with that of the Mg~II system. Assuming that the
scattered light correction for the FOS spectrum is accurate, the $z=0.4415$
system has an optical depth at the Lyman limit of $\tau_{LL} \simeq 1.6$, or
log N(H~I)$\simeq 17.3$. Thus, there is evidence that a significant component
of the extended gas giving rise to the Mg~II absorption has disk-like kinematics
despite the relatively low total H~I column density.   

\subsection{Q1148+387 (4C 38.31)}

The $z_{abs}=0.5531$ absorption system was first identified by Steidel \& Sargent 1992 (SS92).
As evident in Figure 3a, the absorbing galaxy is a mid-type spiral only moderately inclined with respect to
the line of sight ($\theta_{\rm i} \simeq 40^{\circ}$), and as such the amplitude of the observed
rotation curve is only $\rm v_{c} sin\theta_i \simeq 125$ \kms. The (somewhat uncertain) corrected
rotational velocity is $\rm v_{c} \simeq 195$ \kms, consistent with or slightly
low given the galaxy's luminosity. The QSO sightline, which is only 14.4h$^{-1}$ kpc in projection from
the galaxy, still apparently samples only blue-shifted gas-phase velocities. However,
the kinematic extent of the gas is significantly broader than the range of velocities
seen from the galaxy rotation curve, and the absorbing gas appears to be distributed
into $\sim 6$ individual velocity components, all with about the same relative
strength of Mg~II and Fe~II $\lambda 2600$ absorption. We discuss kinematic models
for this system in \S 4.3.

The total H~I column density of this system can be estimated from an archival FOS
spectrum. Again assuming that the zero point of the FOS flux scale is accurate,
the optical depth is $\tau_{LL} \simeq 1.5$, or log N(H~I)$\simeq 17.2$. This would
appear to be an unexpectedly low H~I column given that the impact parameter is at a
projected galactocentric distance of only $14.4h^{-1}$ kpc (corresponding to
a disk impact parameter of $18.8h^{-1}$ kpc given the inclination angle). Apparently either the
QSO sightline has found a ``hole'' in the H~I distribution, or the outer disk of
G1 1148+387 is relatively H~I--poor. 

\subsection{Q1222+228 (Ton 1530)}

The originally-targeted Mg~II redshift along this line of sight was the $z_{abs}=0.6681$
system first discovered by Young, Sargent, \& Boksenberg (1982). We believed,
on the basis of galaxy spectra of admittedly marginal quality, that this system
was produced by the nearly-edge-on spiral which was the primary target of our LRIS
spectroscopy. We were somewhat surprised to find that the galaxy instead
has an emission redshift of $z_{gal}=0.5502$ (leaving the much stronger
$z_{abs}=0.6681$ absorption system without a confirmed absorbing galaxy candidate)
from the much-higher-quality Keck/LRIS spectra. An {\it a posteriori} search
of the HIRES spectrum yielded a weak Mg~II doublet, with
$W_0(\lambda 2796) = 0.08$ \AA, so that it is well below the detection threshold
of the SS92 survey, and most earlier Mg~II absorption line surveys.  We later realized 
that this weak system was cataloged by Churchill \et 1999, at $z_{abs}=0.55020$. 

Ignoring for the moment the absence of an identified absorbing galaxy for
the $z_{abs}=0.6681$ system (there are several candidates without spectroscopic
redshifts, albeit at relatively
large impact parameters, evident in Figure 4a), the $z_{abs}=0.5502$ system is interesting
for a number of reasons: first, the absorption is apparently associated
with a much fainter galaxy than the other 4 systems considered in this 
pilot study, with a circular velocity of only $v_c \simeq 100$ \kms. The
galaxy is highly inclined ($\theta_{\rm i} \simeq 75^{\circ}$), with the projection of
the major axis missing the QSO position by only 1.5 arc seconds on the plane
of the sky. Despite the high inclination and the projected impact parameter of 26.5h$^{-1}$ kpc,
the absorption velocity is consistent with the systemic
velocity of the galaxy (see Figure 4b), and has a total velocity spread of only
$\sim 50$ \kms.  The implications of the relative kinematics of the absorption
and emission will be discussed in \S 4.4.

There is no information on the H~I content of either the $z_{abs}=0.5502$ or the
$z_{abs}=0.6681$ system, despite the existence of HST/FOS spectra, as the
continuum of the QSO is cut off shortward of $\sim 2200$ \AA\ by a higher
redshift Lyman limit system. 

\subsection{Q1317+276 (Ton 153)}

SS92 discovered two intermediate redshift absorption systems along this line of sight,
at $z_{abs}=0.2887$ and at $z_{abs}=0.6598$. An object within about 1 arc second of the
QSO sightline is evident in the {\it HST} image presented in Figure 5a, but no
successful spectroscopy has been performed on this object, since the QSO is extremely
bright. Earlier spectroscopy from Lick Observatory indicated that both
G1 and G2 were consistent with having the same redshift as the $z_{abs}=0.6598$
absorption, within the errors (the spectra were of low quality). Because of
this ambiguity, the LRIS slit was oriented so as to include both galaxies, and
hence is not aligned with the major axis of either galaxy. The new spectra clearly
show that G1 (which is both spectroscopically and morphologically of early type)
has $z_{gal}=0.672$, which is much too high to be related to the observed absorption.
Galaxy G2, on the other hand, has a systemic redshift of $z_{gal}=0.6610$, within
200 \kms of the observed Mg~II absorption (see Figure 5b). 

Galaxy G2 has a very large projected impact parameter (71.6$h^{-1}$ kpc) and because
of this the identification of it as the galaxy responsible for the absorption
is somewhat tentative; however, as for other more secure identifications presented
in this paper, the extrema of the absorbing gas velocities are consistent with an
extrapolation of the disk kinematics to the large galactocentric distances. In this
case, though, attributing the gas to the disk of G2 would 
would imply a de-projected circular velocity of $\simeq 250$ \kms (which is quite
reasonable for a galaxy of $\sim 1.6L^{\ast}$) but implies an extension of
the disk--like rotation to a de-projected galactocentric radius of $\sim 130$h$^{-1}$ kpc,
possibly stretching the bounds of feasibility\footnote{The circular velocity implied
for G2 is also consistent with the measured rotation curve, although the fact
that the slit was placed at an angle of $\sim 60$ degrees relative to the major axis
makes the de-projection of the rotation curve somewhat model-dependent.}.   

The absorbing gas yields a clear detection of Mg~I associated with the dominant component
of Mg~II at $v_{sys}=-170$ \kms, and a hint of Mg~I in the weaker $v_{sys}=-80$ \kms component. 
There does appear to be significant Mg~II absorption all the way to 
$v_{sys}=0$. This is circumstantial evidence that the gas is indeed dynamically related
to galaxy G2. We will discuss this system in more detail in \S 4.5. 

There is a strong Lyman $\alpha$ absorption line ($W_{\lambda}^0=1.5$ \AA) associated with the Mg~II
absorption, and an optically thick Lyman limit (measured by Bahcall \et 1993 at
$\tau = 3.5$ and Churchill \et 2000
at $\tau = 5.4$. Given uncertainties in the flux zero point of FOS data,
we take these measurements as lower limits, suggesting an H~I
column density of $\simgt 10^{18}$ cm$^{-2}$). 
Interestingly, there is
a strong complex of Lyman $\alpha$ forest lines extending from $z=0.660$ to $z=0.672$
($\sim 2200$ \kms; Bahcall \et 1996) and there are at least 2 galaxies (including the early type
G1 1317+276) near the line of sight at $z \simeq 0.67$. 

\section{KINEMATICS}

Although the present sample of galaxies is small, there are interesting trends
that already suggest that standard interpretation of Mg~II kinematics may be inadequate
to describe the data.  It has become standard to interpret the velocity asymmetries
that are typical of Mg~II absorption systems (and DLAs) as being consistent with the interception
of a rotating disk that is highly inclined with respect to the line of sight (LOS) 
(e.g., Lanzetta \& Bowen 1992; Charlton \& 
Churchill 1998; Prochaska \& Wolfe 1997). In the sense that most of the 5 absorbing
galaxies observed were selected because they are clearly disk galaxies, it is perhaps
not surprising that the kinematics are indeed similar to that expected for
disk rotation. Charlton \& Churchill (1998) have interpreted many Mg~II systems
(drawn primarily from the same sample as the present observations) as having a dominant component
ascribed to rotation, with additional absorption due to radial infall (i.e., a ``halo''
component) to explain the weaker ``satellite'' absorbing components. 
Components of Mg~II absorption ascribed to radial infall or outflow would be
expected, on average, to be roughly symmetric with respect to the galaxy
systemic redshift. The same would be true if there were a significant component
of ``halo'' gas clouds in random orbits about the galaxy center of mass (see, e.g., Charlton
\& Churchill 1998). 

However,
the new information on the systemic velocity and rotation curves for the
parent galaxies presents a puzzle:  
for 4 of the 5 systems (i.e., all but Q1222+228), {\it all} of the absorbing material
lies at velocities offset to one side of the galaxy systemic redshift $z_{\rm gal}$.
Asymmetric behavior of the absorption line kinematics with respect to galaxy systemic
velocities would be a clear prediction of gas clouds embedded in a rotating disk
with relatively  
large inclination angle with respect to the line of sight (cf. Lanzetta \& Bowen 1992). 
However, at the large
observed impact parameters to the absorbing galaxies, it is not possible to explain
the {\it range} of velocities $\Delta v$ spanned by the absorbing material with a thin
or even a moderately thick disk model. At large
impact parameter, for a highly inclined rotating disk, the sampling of the galaxy rotation
curve would be such that only a very small $\Delta v$, centered at $v \simeq v_{\rm rot}$,
would be expected. There are velocity components that would be consistent with an extension
of disk kinematics to the line of sight in each case, but reproducing the {\it width} of the
velocity profiles, given our knowledge of the galaxy geometry, requires an additional component to the kinematic model.

To summarize, the gas-phase kinematics traced by Mg~II absorption relative to the
galaxy systemic redshift and rotation curves seems to indicate that rotation
dominates, but simple disk rotation is inadequate to explain the range of
velocities. The fact that even the sub-dominant components of absorption 
exhibit systematics that are also indicative of a preferred axis of rotation,
as opposed to a symmetric distribution of velocities relative to systemic, suggests that
absorbing gas has net rotation independent of its location within the galaxy's gaseous envelope.  
We now consider a simple kinematic model that allows for rotation-like systematics with 
the possibility of extending the velocity range of absorbing gas without introducing
velocity components on the opposite side of the systemic galaxy redshift $z_{\rm gal}$.
These models are similar to those presented by Morton \& Blades (1986) to explain the
kinematics of Ca~II absorption toward stars in the Galactic halo. 

%\subsection{Kinematic Models}

We assume a co-rotating thick disk (as has often been invoked in
studies of high ionization species in the Galactic halo--see, e.g., Savage, Sembach, \& Lu 1997), 
where the effective thickness $H_{eff}$ in
the $z$ direction is a free parameter,   
and a flat rotation curve throughout. The non-negligible thickness allows
for a larger range of velocities to be sampled by a line of sight intersecting
it, as long as the inclination angle $\theta_{\rm i}$ with respect to the plane of
the sky is non-zero.  The assumed rotation curve is given by:
\begin{equation}
\rm
    \vec V_{\phi}(r) = v_{c} \hat {\phi}
\end{equation}
where $r$ is a radial coordinate measured in the plane of the galaxy disk and $\hat{\phi}$
is the tangential direction. 

\noindent
In Cartesian coordinates:
\begin{equation}
\rm
   \vec v = {\frac{-y}{r}}v_{c} \hat{\it x}+{\frac{x}{r}}v_{c} \hat{\it y} 
\end{equation}
The coordinates are defined so that the value of $x$ is constant 
along the line of sight, $x = p$, where $p$ is the impact parameter measured
along the major axis of the galaxy (see Figure 6). The line of sight is
parallel to the $y$-$z$ plane, as illustrated in Figure 6. \\

We describe the line of sight by a vector of the form:
\begin{equation}
\rm
 \vec {\bf L}  =  -sin{{\theta_i}} \hat{\it y} - cos{{\theta_i}} \hat{\it z} 
\end{equation}
where $z$ is the coordinate measured perpendicular to the disk plane of the galaxy.
Projecting the velocity vector above along the LOS vector yields:

\begin{equation}
\rm
 v_{los}  =  \frac{-v_{c}sin{\theta_i}} {\sqrt{1+(\frac{y}{p})^{2}}}
\end{equation}

We make the model more general by introducing a z--dependence of the rotational
speed at a given radial distance from the disk rotation axis.  
The model is parameterized by a simple exponentially
declining velocity with an adjustable velocity scale height.  
\begin{equation}
\rm 
    v_{\phi}(r,z)  =  v_{c} e^{- (|z|/h_{v})} \hat {\phi} 
\end{equation}

\noindent
where ${\rm h_{v}}$ is the velocity scale height, ${\rm v_{c}}$ is the measured
circular velocity at mid-plane, and $z$ is the height above the mid-plane of the disk. 
The line of sight velocity as a function of $y$ is then given by:

\begin{equation}
\rm 
 v_{los}  =  {\frac{ - v_{c}sin{\theta_i} }
{\sqrt{1+(\frac{y}{p})^{2}}}} e^{-{(|y-y_{0}|}/(tan{\theta_i}h_{v}))} 
\end{equation}
where the minus sign indicates the sense of rotation. 
The parameter $y_0$ is the $y$ value of the intersection of the line of sight
with the mid-plane of the disk (see the diagram in Figure 6). The range of $y$ values
encountered as the line of sight pierces the gas distribution surrounding the
galaxy then ranges from ${\rm y_0-H_{eff}tan \theta_i}$ to ${\rm y_0 +H_{eff} tan \theta_i}$. In terms
of the $y$ coordinate, the distance along
the line of sight, relative to the point where it intersects the projection of
the disk mid-plane, is then just ${\rm D_{los}=(y-y_0)/sin \theta_i}$.  

With direct measurements of $\theta_i$, p, ${\rm y_0 cos{\theta_i}}$ from the HST images, and
the circular velocity ${\rm v_{c}}$ from the optical rotation curves
(all measured values are summarized in
Table 5), the only free parameters in
these simple models are the assumed disk thickness ${\rm H_{\rm eff}}$ which controls the
range of $y$ values considered, and the velocity scale height ${\rm h_v}$. 

Our approach in fitting the models was to adjust the two parameters ${\rm H_{eff}}$
and ${\rm h_v}$ to try to
match the observed gas-phase kinematics. We have made no assumptions about
the distribution of Mg~II absorbing clouds as a function of $z$ distance or
$r$ with respect to the center of each galaxy. Instead, for simplicity, we treat $H_{\rm eff}$ 
as the effective thickness of the gas layer capable of giving rise to
detectable Mg~II absorption and so we neglect the path length differences through
various parts of the gas layer. In some cases there are many combinations of
${\rm h_v}$ and ${\rm H_{\rm eff}}$ that are in reasonable agreement with the observed velocities
of the absorbing gas, while in other cases the allowed range of parameters is very
small. We discuss each case individually below, and illustrate particular models that
come closest to agreement with the data in Figure 7. 
The adopted parameters
that were used to produce the plots shown in figure 7 are summarized in table 6.

\subsection{Q0827+243}

The large velocity range ($\simeq 270$ \kms) observed in the absorption profile
of G1 0827+243 cannot be reproduced by any thick disk model without incorporating
a declining rotation speed as a function of $z$ distance. The model shown
in figure 7a, which accounts for most of the observed velocity of the absorbing
gas, is not unique-- a similar velocity range can be produced by any model
for which $h_v$ is small compared to $H_{\rm eff}$ (i.e., where there is gas
with essentially all velocities between $-250$ \kms and $20$ \kms with respect
to systemic along the line of sight) and the effective thickness of the gas
layer $H_{\rm eff} \ge 5h^{-1}$ kpc.  

The fact that the maximum velocity in absorption exceeds the projected rotation curve
may mean either that the rotation speed of the galaxy has been under-estimated,
or that there is a turbulent component of velocity which adds $\sim 50$ \kms
to the gas-phase near the intersection with the disk. As discussed above, 
the galaxy is apparently being disturbed by a smaller satellite galaxy, and
the galaxy is by far the most actively star forming, judging by the equivalent
width of the [OII] $\lambda 3727$ emission line, in the present sample. Both of
these phenomena may contribute to the gas-phase kinematics.   

\subsection{Q1038+064}

The velocity profile of G1 1038+064 is well-reproduced by models in which 
$h_v$ is of the same order as $H_{\rm eff}$ and $H_{\rm eff} \ge 2h^{-1}$ kpc. 
Gas with the largest departure in line-of-sight velocity relative to systemic
is predicted by this family of models to lie
in the plane of the galaxy, so that it is quite reasonable to use the
absorption line velocities to extend the galaxy rotation curve to $45h^{-1}$
kpc as discussed in \S 3.2. Nevertheless, it is somewhat surprising to observe
disk kinematics in gas with such small H~I column density (see \S 3.2).

\subsection{Q1148+387}

It is difficult to reproduce the full velocity range of $\Delta v \simeq 160$ \kms
with our kinematic model. The most successful model is one in which the effective
gas layer is very thick ($\simgt 25h^{-1}$ kpc) and has a constant rotational
velocity with $z$ distance (as in figure 7c). With this model, the extreme value
of the line of sight velocity reaches only $-120$ \kms with respect to systemic
(although the extension to zero systemic velocity is reproduced) and most of the
absorption occurs substantially away from the plane of the disk, which may
be unphysical. Models with
relatively thin gas layers and small velocity scale heights, which work for the
two previous cases, are very unsuccessful for G1 1148+387.   
It is possible that models in which the angular momentum vector for ``halo'' gas 
has a different direction than that of the
disk would be more successful. We merely point out that application of our
simple kinematic model leads to a relatively poor agreement with the observations
and a somewhat implausible physical situation. 

\subsection{Q1222+228}

The difficulty for this system is that the galaxy is nearly edge-on and yet
the absorption kinematics are centered at the systemic velocity of the galaxy.
Thus, it is necessary to keep the extension of the disk which carries the full
rotation speed of the galaxy very thin to prevent absorption components with high
negative velocities. At the same time, the  total thickness of the gas layer,if
associated with a thick disk, must be large enough to allow for absorbing gas
$\sim 5$ kpc above the projection of the disk plane. 
Reasonably successful models are those with $H_{\rm eff} \le 5h^{-1}$
kpc and velocity scale heights that are very small, $h_v \le 0.1h^{-1}$ kpc. 
Such a kinematic model produces gas that asymptotes quickly to zero systemic velocity
only a few kpc above the disk plane.

It is notable that the Mg~II absorption for this system is much weaker, and
the galaxy considerably fainter, than for
the others considered here.

\subsection{Q1317+276}
 
Reproducing the large departure of the measured absorbing galaxy velocities 
from systemic requires gas at large $z$ distances above the plane to
have very close to the full circular velocity of the galaxy. Because 
the line of sight through the galaxy is nearly parallel to the galaxy minor axis, 
material in the plane of the disk cannot contribute to the observed
velocities. For the model shown in figure 7e, most of the absorption
at non-zero systemic redshifts arises in material very nearly directly above the center of the
galaxy rotation axis (i.e., very small values of $y$--see figure 6) which is rather
unphysical (recall that we have assumed a constant rotational velocity, which is acceptable
at large galactocentric distances).  

As discussed in \S 3.5, the identification of galaxy G2 1317+276 as the one responsible
for the absorption must be considered tentative. Nevertheless, 
there is gas at velocities that range from systemic to a maximum velocity that
is very similar to the maximum rotation speed that could be seen associated with G2 1317+276
given the inclination angle. This system would seem to be one in which domination by ``halo''
velocity components might be the most reasonable, but once again the absence of
gas at positive velocities with respect to systemic is puzzling. 

\section{DISCUSSION}

We have presented data on 5 intermediate redshift Mg~II absorption systems
for which there is much more information than has been available in the past,
and in many ways it has made for a more puzzling picture of the nature
of absorbing material in the outer parts of (in these cases, disk) galaxies. 
Most recent models for the kinematics of Mg~II absorption systems have involved
two separate components contributing to the kinematics-- a rotating disk component,
which is thought to produce the ``dominant'' components of complex kinematic
systems (e.g. Charlton and Churchill 1998, Churchill \& Vogt 2001), with more
symmetrical (with respect to the systemic redshift) ``halo'' components providing
a broad distribution of velocity, whose origins might be ascribed to random motions,
infall, or outflow. Given the limited geometrical information from Mg~II--selected
galaxy surveys (Steidel \et 1997, SDP, Bergeron \& Boiss\'e\ 1991)
at these redshifts, and the limited morphological information on the absorbing
galaxies (Steidel 1998; Steidel \et 1997), this picture has the advantage of
being compatible with both the spiral nature of most of the galaxies, and
the large (and possibly quasi-spherical) gaseous envelopes surrounding a very
large fraction of $z \sim 0.6$ field galaxies within $\sim 1.5$ magnitudes of
present-day $L^{\ast}$. 

However, based on the systems presented in this paper, the situation cannot be so simple--
the components that are not easily explained by thin disk rotation must have
the systematics that are like those produced by rotation. 
There is clear evidence for rotation in 4 out of the 5 cases, since not only 
is all of the absorption offset to one side of the systemic redshift, in all
cases it is in the right sense to be qualitatively explained by an extension 
of the disk rotation to the line of sight (which is well beyond where the optical
rotation curves are measured, by factor of 2 to as much as a factor of 6). 
As detailed in \S 4, though, given knowledge of the disk inclinations and
impact parameters (not known for previous analyses of Mg~II kinematics), 
disk--like rotation is not enough to explain the bulk of kinematic components
seen in absorption.  Our simple kinematic modeling of rotating ``thick disks''
in \S 4 (which are by no means intended to be unique solutions to the kinematic
conundrum) imply that for two of the galaxies (G1 1148+3842 and G2 1317+276)
a successful model does not involve an extension of the disk at all, but
requires essentially a {\it rotating halo}. In the three other cases,
the observed kinematics are reasonably well fit by thick disk models
in which the circular velocities are a rapidly decreasing function of
scale height, with the extrema of the velocities being produced at the
disk intersection, but where the gas at lower relative velocities with respect to
systemic comes from fairly
large $z$ distances. As discussed in \S 4, the galaxies best--fit
with the velocity scale height models allow for a wide range of
solutions so long as the ratio $H_{\rm eff}/h_v$ remains roughly constant,
as there are only weak constraints on the required thickness of
the gas layer above the extrapolation of the plane of the disk.   
Given what is known about the geometry of the gaseous envelopes capable
of giving rise to easily detectable Mg~II absorption, we would be inclined
to favor models in which $H_{\rm eff}$ is on the order of $\sim 30-40h^{-1}$
kpc for $L^{*}$ galaxies like G1 1038+064 and G1 0827+243. 
On the other hand, G2 1222+228,
which is among the faintest Mg~II absorbers identified at comparable redshift,
must be quite different from these larger spirals, in that the effective thickness
of the disk must be quite thin to {\it avoid} producing absorption at large
velocities with respect to systemic, and the velocity scale height must be
even smaller to bring the kinematic model into tolerable agreement with the data. 

It is instructive to consider possible local analogs of the kinematic behavior
of galactic gas we observe at $z \sim 0.5$. For highly ionized gas in the Galaxy
seen in absorption against the continua of hot stars and extragalactic AGN,
a model in which the ``halo'' gas co-rotates with the disk up to heights of
several kpc above the plane has often been assumed, and recent observations
seem to require this (Savage, Sembach, \& Lu 1997).  The scale heights reached
appear to vary as a function of the ionization state of the ion, with the
most highly-ionized species extending to the largest distances above
the plane. However, in at least 2 of the cases we have observed, Mg~II would have to
have an effective scale height that is more than an order of magnitude larger 
than that of C~IV in the Galaxy, and in the other cases a completely co-rotating halo would fail 
badly to reproduce the observed absorption line kinematics. 

A recent development based on
the most sensitive H~I (e.g., Swaters, Sancisi, \& van der Hulst 1997; 
Sancisi \et 2000, Schaap, Sancisi, \& Swaters 2000) and H~II (e.g., Rand 2000)
measurements locally is the ability to follow the kinematics of the gas to relatively large
$z$ distances above the planes of spiral galaxies. Rand (2000) finds that
for the edge-on starburst galaxy NGC 5775, there is gas whose rotation velocity has
decreased to zero by a height of 5 kpc. He interprets this behavior as 
a trend of decreasing rotational velocity as a function of $z$, similar to
our modeling above. Swaters \et (1997) observed clear evidence for 
a systematically smaller rotation velocity of the ``H~I halo'' of NGC 891, with
gas at several kpc above the plane rotating 25--50 \kms more slowly than in
the plane. Sancisi \et (2000) discuss 21-cm observations of galaxies with ``beards'',
in which gas observed away from the plane of the galaxy seems to ``know''
about the rotation of the disk (i.e., the rotation has the same general direction as far as
can be discerned with the observations) but has kinematics that represent large
departures from the disk rotation. The qualitative similarities of these
observations to those presented in this paper are clear; however, it is
unclear how common the kinematically ``anomalous'' gas is in local spirals
(only a few galaxies have been observed to the required level of sensitivity). In any
case, the gas-phase kinematics of $z \sim 0.5$ galaxies refer to much larger galactocentric
distances and much smaller H~I column densities (with the exception of G1 0827+243
which would easily be observed in 21 cm emission if it were nearby). 

The interpretation of the slowly rotating gas in nearby galaxies is largely
qualitative at the present time-- the authors cited above invoke both
hydrodynamic disk/halo cycling of gas, and changes in the gravitational
potential with $z$ distance above the plane, as possible explanations of the observations. If the
higher redshift objects are at all analogous, it is hard to imagine that
the gravitational potential argument can be relevant, since the sight lines all
intersect the galaxies at radii where dark matter would be expected to
dominate strongly over a baryonic disk. 

Bregman (1980) considered kinematic models of the disk/halo circulation (the
``Galactic fountain'' model) in which
parcels of hot gas are expelled from the disk by star formation events and may, by the
time they cool, have been transported to both a large height above the plane and
to a larger distance from the galaxy rotation axis. Because of conservation of
angular momentum, the gas at large heights above the galactic plane would
lag with respect to the disk rotation, producing a more slowly-rotating ``halo''.
In practice, the kinematics of the gas as a function of height $z$ would depend
on the details of the distribution and energetics of previous star-formation episodes
and on the pressure profile of the galaxy as a function of galactocentric distance; 
in principle, gas with any velocity between systemic and $v_{c} sin\theta_{\rm i}$ could be observed
at any position along the line of sight. Nevertheless, as long as the rotation of the
gas dominates over radial motions (infall or outflow) such a picture would be
qualitatively consistent with our observations.  In the nearby starburst galaxies
discussed above, very active current star formation lends qualitative credence to
fountain flows as a possible explanation of the observed gas-phase kinematics, but
this leads to a puzzle for the higher redshift objects-- with the exception of
G1 0827+243, none of the galaxies considered here has an unusually high rate of
star formation, and certainly there is no active star formation coinciding with
the large inferred disk impact parameters. The observation that the Mg~II
absorbers at intermediate redshift tend in general {\it not} to be particularly active
star-formers has been used previously to argue against fountain-type flows 
being important to the presence of extended gaseous envelopes (SDP, Steidel 1995), 
but this argument assumes
that the timescale for the circulation of the gas is relatively short. We now consider
the possibility that past star formation, now observed only through the older
stellar populations present in the galaxies, might be responsible for the rotating
halo gas observed at $z \sim 0.5$.

There is mounting evidence for the importance of large-scale galactic winds for 
star forming galaxies at high redshift. Observations of $z \simgt 3$ Lyman break
galaxies show clearly that the strong far-UV interstellar absorption lines are blue-shifted,
and the Lyman $\alpha$ emission lines red-shifted, by up to 1000 \kms with respect
to systemic (Pettini \et 1998, 2001), with typical implied outflow velocities being
several hundred \kms. Very recently, it has been shown that the properties of the
Lyman $\alpha$ forest are strongly affected by the presence of LBGs $z \sim 3$, and that several
different observations can be explained simultaneously if the super-winds
have a sphere of influence of $\sim 125h^{-1}$ kpc on average (Adelberger \et 2001). The cooling time for
the shock heated gas in the halo can be very long, and it is at least conceivable that 
this gas, or similar gas ejected at lower redshifts (where there are currently fewer observations
constraining the extent of super-winds), could ultimately supply the gas that forms 
the bulk of the disk observed at $z \sim 0.5$ and the material that produces Mg~II
absorption. The physics of the multi-phase gas that no doubt results from the wind activity
is very complex, and a full treatment is well beyond the scope of this paper. It is not
clear that this kind of flow would really result in a ``memory'' of disk-like kinematics
in the halo, since the material involved in the type of super-winds inferred to exist
at $z \sim 3$ would tend to originate from low angular momentum gas that has settled
to a very compact nuclear region where most of the star formation appears to take place.
An alternative possibility is that the gas has acquired significant angular momentum during
the extended time that it spends at large galactocentric distances, and that this angular
momentum is naturally strongly correlated with that of the gas that has found its way to
the disk. In this scenario, much of the gas
falling onto the disk would do so gradually, would be significantly metal-enriched, and would have
the same kinematic systematics as disk gas, albeit with smaller rotational velocities. It is not
entirely clear whether the observed kinematics at $z \sim 0.5$ are consistent with both
rotation {\it and} infall; the  sample is too small to justify a more in-depth treatment
at this time. 
In any case, a general picture of halo gas being ``recycled'' disk gas may actually help explain the loose correlation of inferred
size of the Mg~II-absorbing envelope with stellar mass, the inferred roughly axisymmetric
geometry of the envelope, and the persistence of absorbing gas over much longer than
the typical galaxy dynamical time of $\sim$ a few $\times 10^8$ years.

It is almost certainly premature to generalize about the nature of the Mg~II absorbing gas, given
the small sample of 5 systems presented here and the fact that each system requires somewhat
different assumptions to find kinematic models that are adequate. We have not considered in
detail whether any of these ad hoc kinematic models are physically plausible, and we
have not considered the hydrodynamics of the gas at all. More detailed modeling seems unjustified
until a larger sample is in hand.  It is clearly worth extending
this type of study in two ways. First, it is essential to obtain accurate redshifts and (where
possible) rotation curves for a larger sample of Mg~II--selected galaxies (the HIRES and WFPC-2 data are already
on hand for a sample of $\sim 25$ Mg~II absorbers). Secondly, it will be important to obtain
high quality absorption line data extending to higher ionization species (like C~IV) to see if
the rotation signatures are as clear in the highly ionized component as they are for
the gas presently traced by Mg~II. Initial forays in this direction have already been
made by Churchill \et (2000), but the archival FOS data are generally of too-coarse resolution
to compare absorption line kinematics in detail. On the other hand, a fraction of the QSOs in the sample for
which we have WFPC-2 images are bright enough for STIS high dispersion spectroscopy in reasonable
integration times, and this should prove a fruitful line of research in the future. 

\bigskip
\bigskip
We would like to thank 
Kurt Adelberger for help with the observations and for many discussions.  
Useful conversations with Betsy Barton-Gillespie, Liese van Zee, and Jason Prochaska are
gratefully acknowledged. CCS and AES have been supported in part by grants
AST95-96229 and AST-0070773 from the U.S. National Science Foundation and by the David and Lucile
Packard Foundation. Early work on the HST data was supported by grant GO-05984.01-94A
and GO-06577.01-95A from the Space Telescope Science Institute.

\begin{deluxetable}{lcccccc}
\tablewidth{0pc}
%\footnotesize
\tabletypesize{\scriptsize}
\tablecaption{Absorbing Galaxy Properties}
\tablehead{
\colhead{Object} & \colhead{$z_{\rm gal}$}  & \colhead{${\cal R}_{\rm AB}$\tablenotemark{a}} & \colhead{${\cal R}_{\rm AB}-K$} 
& \colhead{$\theta$\tablenotemark{b}} & \colhead{d (kpc)\tablenotemark{c}}  & \colhead{$\rm M_B$\tablenotemark{d}}
} 
\startdata
G1 0827+243 & 0.5258 & 20.83 & 3.88 & ~5.8 & 25.4 & $-$19.98 \\ 
G1 1038+064 & 0.4432 & 20.95 & 4.45 & ~9.7 & 38.8 & $-$19.58 \\ 
G1 1148+387 & 0.5536 & 21.59 & 3.49 & ~3.2 & 14.4 & $-$19.33 \\ 
G1 1222+228 & 0.5502 & 22.82 & 4.02 & ~5.9 & 26.5 & $-$18.09 \\ 
G2 1317+276 & 0.6610 & 21.54 & 3.84 & 14.7 & 71.6 & $-$19.95 \\   
\enddata
\tablenotetext{a}{From ground-based photometry; the ${\cal R}$ filter has an effective
wavelength of 6830 \AA\, close to the HST F702W filter used with WFPC-2. } 
\tablenotetext{b}{Projected angular separation of absorbing galaxy from QSO sightline, in
arc seconds.}
\tablenotetext{c}{Projected proper separation of galaxy from QSO sightline, in $h^{-1}$ kpc, 
for $\Omega_m=0.3$, $\Omega_{\Lambda}=0.7$ cosmology.}
\tablenotetext{d}{Absolute rest-frame B magnitude of galaxy, for $h=1$,$\Omega_m=0.3$,$\Omega_{\Lambda}=0.7$.}
\end{deluxetable}
\begin{deluxetable}{lccc}
\tablewidth{0pc}
%\footnotesize
\tabletypesize{\scriptsize}
\tablecaption{WFPC-2 Observations}
\tablehead{
\colhead{Field} & \colhead{Date (UT)}  & \colhead{Exposure (s)} & \colhead{$\theta_{\rm i}$ (deg)\tablenotemark{a}} 
} 
\startdata
Q0827+243 & 1995 May 29 & 4600 & 69 \\ 
Q1038+064 & 1995 May 31 & 4600 & 60 \\ 
Q1148+387 & 1995 May 31 & 4700 & 40 \\ 
Q1222+228 & 1997 June 11 & 5000 & 75 \\ 
Q1317+276 & 1995 June 01 & 4800 & 58 \\ 
\enddata
\tablenotetext{a}{Galaxy inclination angle with respect to the plane of the sky, estimated from
elliptical isophote fits to the WFPC-2 image of each galaxy}
\end{deluxetable}
\begin{deluxetable}{lccccl}
\tablewidth{0pc}
%\footnotesize
\tabletypesize{\scriptsize}
\tablecaption{LRIS Absorbing Galaxy Spectroscopy}
\tablehead{
\colhead{Object} & \colhead{$z_{\rm gal}$}  & \colhead{Date (UT)} & \colhead{$\lambda$ Range (\AA)} 
& \colhead{Exposure (s)} & \colhead{PA (deg)}
} 
\startdata
G1 0827+243 & 0.5259 & 1999 Mar 18 & 5400--7960 & 2400 & 72.3 (mask) \\ 
G1 1038+064 & 0.4428 & 1999 Mar 19 & 4980--7530 & 2400 & 73.6 (mask) \\ 
G1 1148+387 & 0.5534 & 1999 Mar 19 & 5190--7760 & 2400 & 89.0 (longslit) \\ 
G1 1222+228 & 0.5502 & 1999 Mar 19 & 5450--8020 & 2400 & 34.6 (longslit) \\ 
G2 1317+276 & 0.6606 & 1999 Mar 18 & 5770--8340 & 1200 & 177.7 (longslit) \\ 
\enddata
\end{deluxetable}

\begin{deluxetable}{lcccc}
\tablewidth{0pc}
%\footnotesize
\tabletypesize{\scriptsize}
\tablecaption{HIRES QSO Spectroscopy}
\tablehead{
\colhead{QSO} & \colhead{$z_{\rm em}$}  & \colhead{Date (UT)} & \colhead{$\lambda$ Range (\AA)} & 
\colhead{Exposure (s)}
} 
\startdata
Q0827+243 & 0.909 & 1998 Feb 27 & 3215--5606 & 7200 \\ 
Q1038+064 & 1.270 & 1998 Mar 01 & 3975--6408 & 7200 \\ 
Q1148+387 & 1.303 & 1995 Jan 24 & 3987--6424 & 5400 \\ 
Q1222+228 & 2.040 & 1995 Jan 23 & 3810--6305 & 3600 \\ 
Q1317+276 & 1.022 & 1995 Jan 23 & 3810--6305 & 3600 \\ 
\enddata
\end{deluxetable}

\begin{deluxetable}{lcccc}
\tablewidth{0pc}
\tabletypesize{\scriptsize}
\tablecaption{Measured Model Inputs}
\tablehead{
\colhead{Object} & \colhead{$\theta$}  & 
\colhead{$v_{\rm c}$} 
& \colhead{ p\tablenotemark{a}} & \colhead{$y_{0}$\tablenotemark{b}}\\ 
\colhead{} & \colhead{(deg)} & \colhead{(\kms)} & \colhead{(h$^{-1}$ kpc)} & 
\colhead{(h$^{-1}$ kpc)}}
\startdata
G1 0827+243 & 69 & 260 & 25.3  &  ~~5.6 \\ 
G1 1038+064 & 60 & 260 & 37.4  &  ~~7.9 \\ 
G1 1148+387 & 40 & 195 & ~6.5  &  ~17.2 \\ 
G1 1222+228 & 75 & 100 & 25.0  &  ~~6.0 \\ 
G2 1317+276 & 58 & 250 & ~0.3  & 135.1 \\   
\enddata
\tablenotetext{a}{This is the projected galactocentric distance, measured
along the major axis of the galaxy, at the point tangent to the line of sight
(see \S 4 and Figure 6).} 
\tablenotetext{b}{The $y$ value of the intersection of the line of sight
with the midplane of the galaxy (see \S 4 and Figure 6).}
\end{deluxetable}

\begin{deluxetable}{lcc}
\tablewidth{0pc}
\tabletypesize{\scriptsize}
\tablecaption{Model Galaxy Parameters}
\tablehead{
\colhead{Object} & \colhead{$\rm H_{eff}$ ($\rm h^{-1}$ kpc)} & \colhead{$\rm h_v$ ($\rm h^{-1}$ kpc)}}
\startdata
G1 0827+243 & ~10 & ~5   \\ 
G1 1038+064 & ~~6 & ~5   \\ 
G1 1148+387 & ~50 & ~*   \\ 
G1 1222+228 & ~~6 & ~0.1   \\ 
G2 1317+276 & 100 & ~*   \\ 
\enddata
\tablenotetext{*}{Asterisks indicate cases for which the kinematics are better fit using
$h_v>>H_{\rm eff}$, which is mathematically equivalent to thick disk with constant
rotational velocity.} 
\end{deluxetable}
\vfill

\newpage
\begin{figure}
\figurenum{1}
\plotone{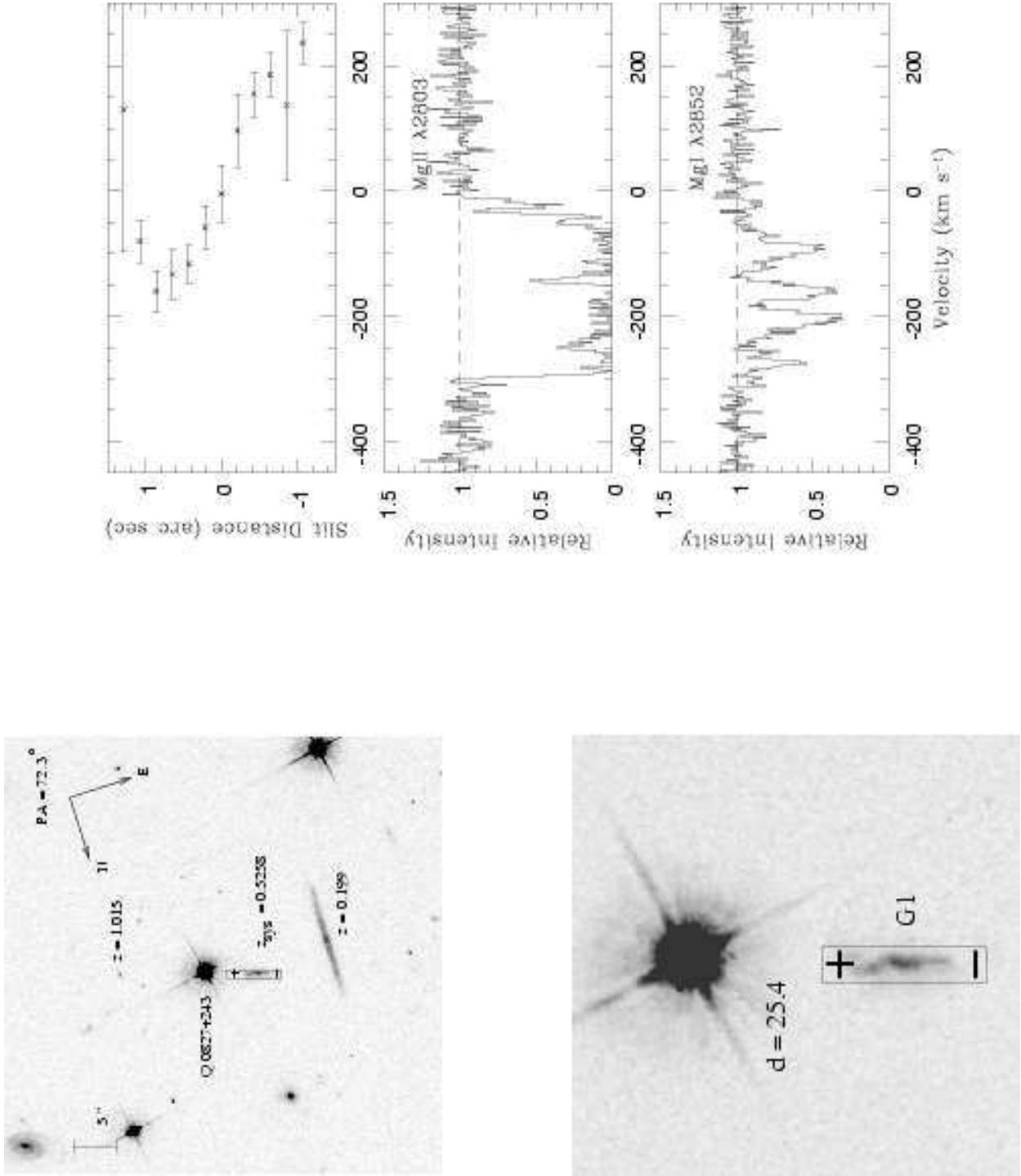}
\caption{a) HST/WFPC-2 F702W images of Q0827+243. The slit position used for
the galaxy spectroscopy is indicated; the relative spatial position along the
slit is indicated with ``$+$'' and ``-'', with ``$+$'' referring to
positive spatial positions, as shown in panel b). b)  
the observed rotation curve (top) of G1 0827+243, and 
the kinematics of the observed absorption (middle, bottom) from the HIRES spectrum
of the QSO.  Note the perturbation
of the galaxy kinematics at positive spatial positions, possibly caused by the
discrepant velocity of the satellite galaxy that is apparent in Figure 1a.
The projected distance ($d$) in $h^{-1}$ kpc between the QSO sightline and
the absorbing galaxy is also indicated. 
 }
\end{figure}
\begin{figure}
\figurenum{2}
\plotone{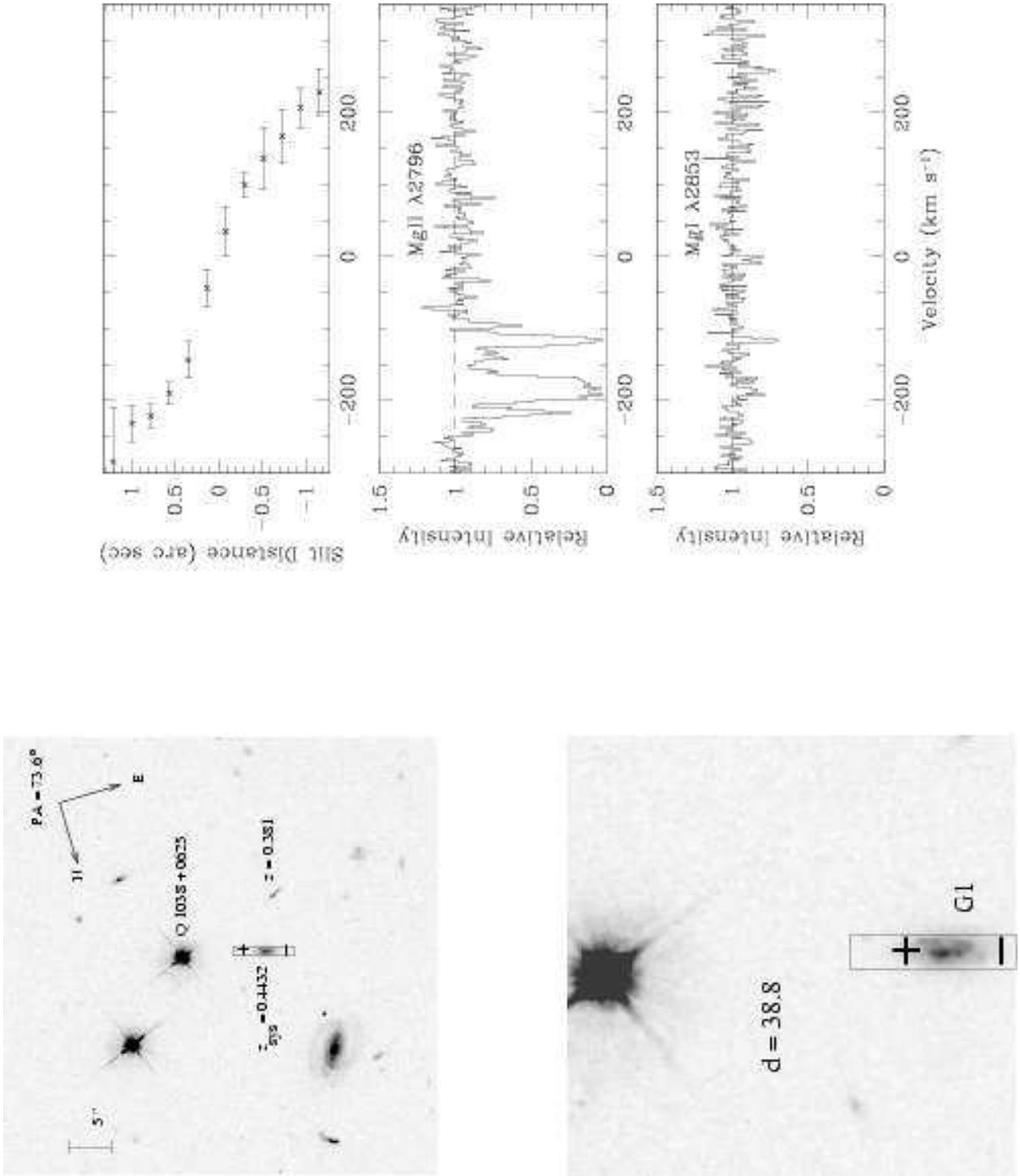}
\caption{a) Same as figure 1, for Q1038+064 b) the observed rotation curve (top)
of G1 1038+064, and the kinematics of the absorption (middle, bottom) from the
HIRES spectrum of the QSO.}
\end{figure}
\begin{figure}
\figurenum{3}
\plotone{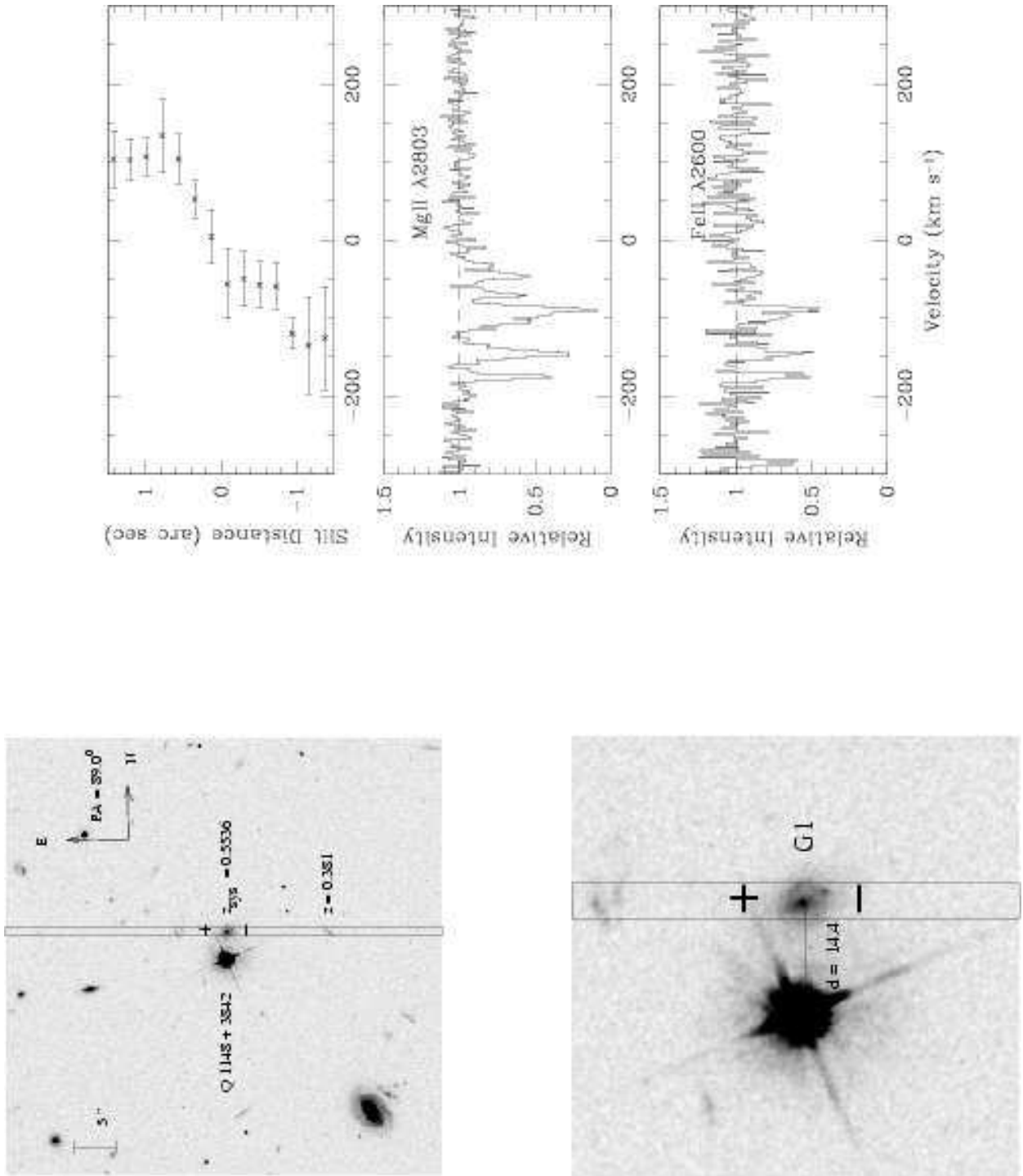}
\caption{a) Same as figure 1, for Q1148+387 b) 
the observed rotation curve (top) of the G1 1148+3842, and
the kinematics of the absorption (middle, bottom) from the HIRES spectrum of the QSO. }
\end{figure}
\begin{figure}
\figurenum{4}
\plotone{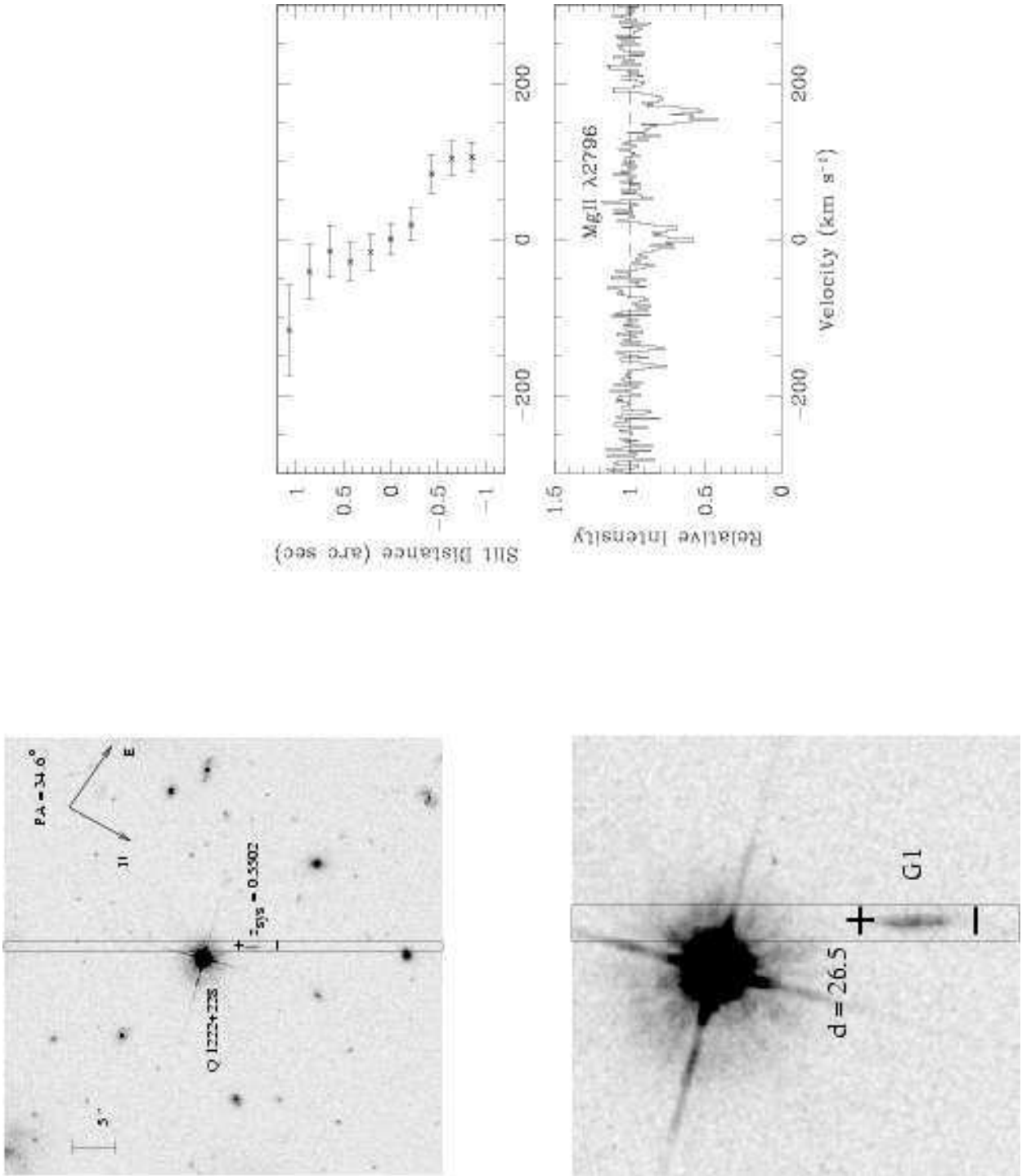}
\caption{a) Same as figure 1, for Q1222+228. b)
the observed rotation curve (top) of the G1 1222+228, and
the kinematics of the absorption (middle) from the HIRES spectrum of the QSO. 
The feature that appears in the
middle panel near $+150$ \kms is from a different redshift system and so should
be ignored.}
\end{figure}
\begin{figure}
\figurenum{5}
\plotone{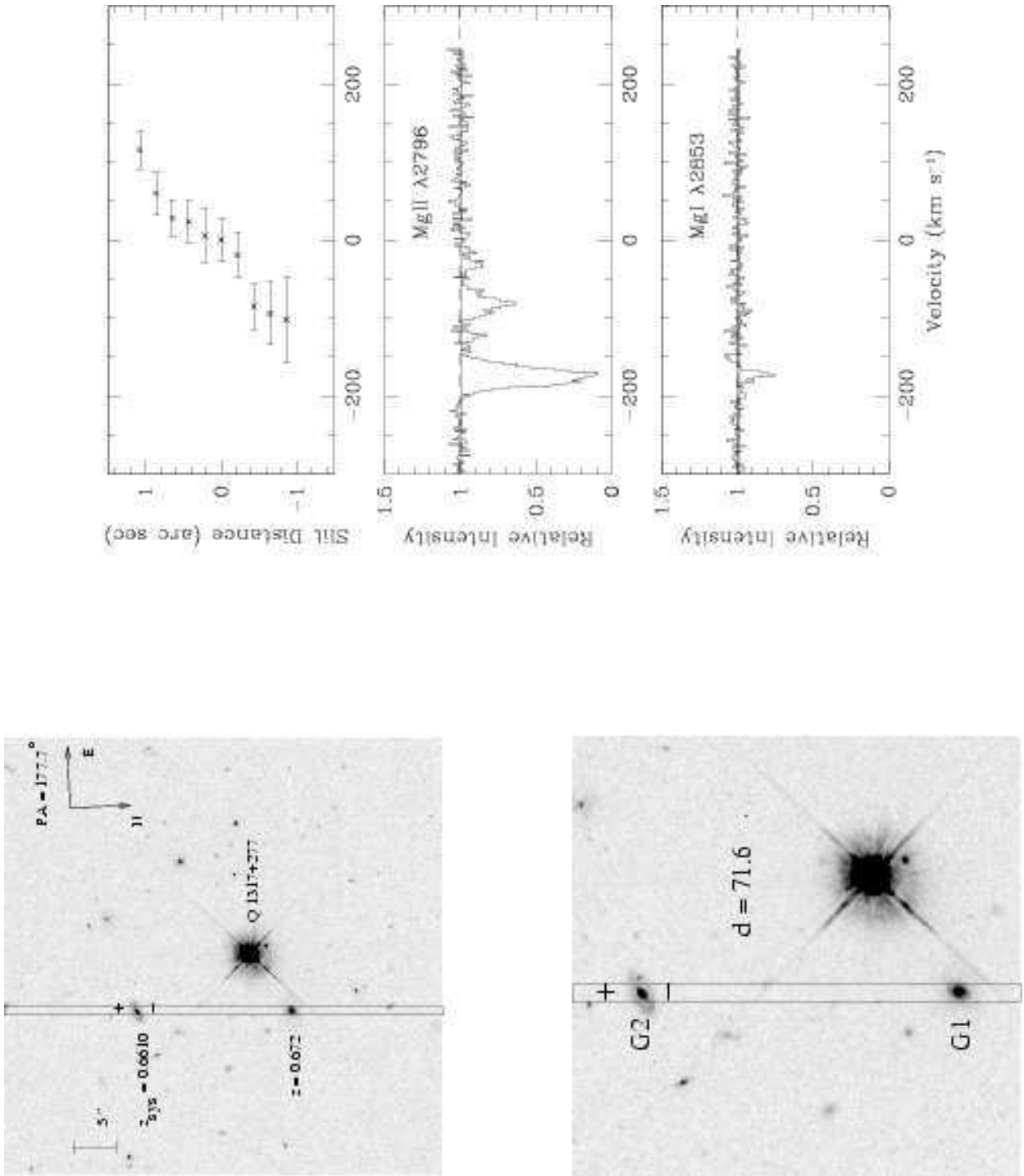}
\caption{a) Same as figure 1, for Q1317+276. b)
the observed rotation curve (top) of G2 1317+276, and
the kinematics of the Mg~II and Mg~I absorption (middle, bottom) from the
HIRES spectrum of the QSO. }
\end{figure}
\figurenum{6}
\begin{figure}
\plotone{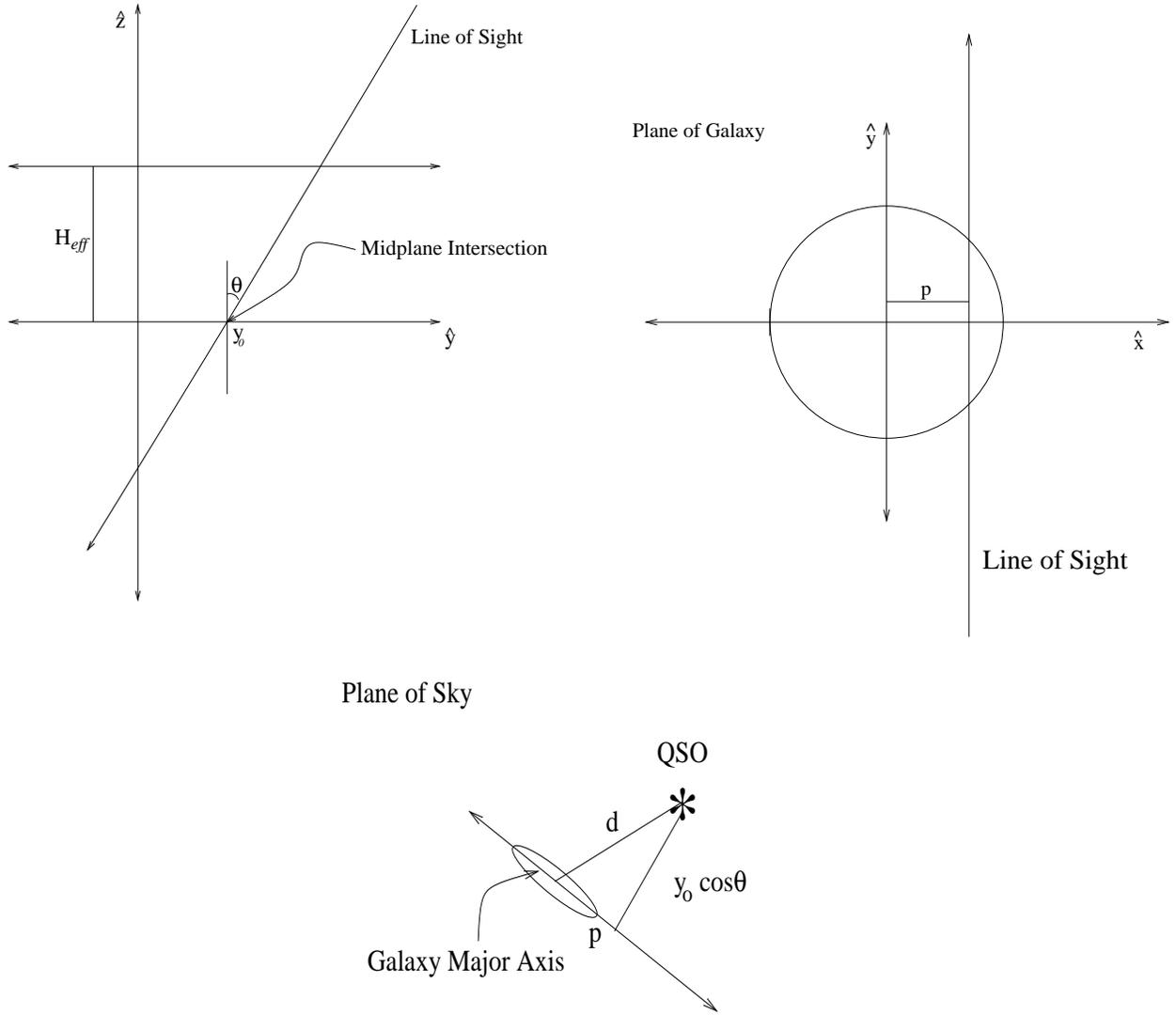}
\caption{Diagrams illustrating the coordinate system used for the
kinematic models described in the text.   
The galaxy disk is located in the $(x,y)$ plane; the axes are defined such
that the line of sight maintains a constant $x$ distance $p$ from the
$y$ axis, and the $y$ axis is parallel to the line of sight.  
$H_{\rm eff}$ is the effective thickness of the
disk of material capable of giving rise to detectable Mg~II absorption, $y_0$
is the galaxy $y$ coordinate where the line of sight intersects the mid-plane. 
$d$ is the projected distance of the QSO sightline from the galaxy centroid
(as in Figures 1a-5a).}
\end{figure} 
\begin{figure}
\figurenum{7}
\plotone{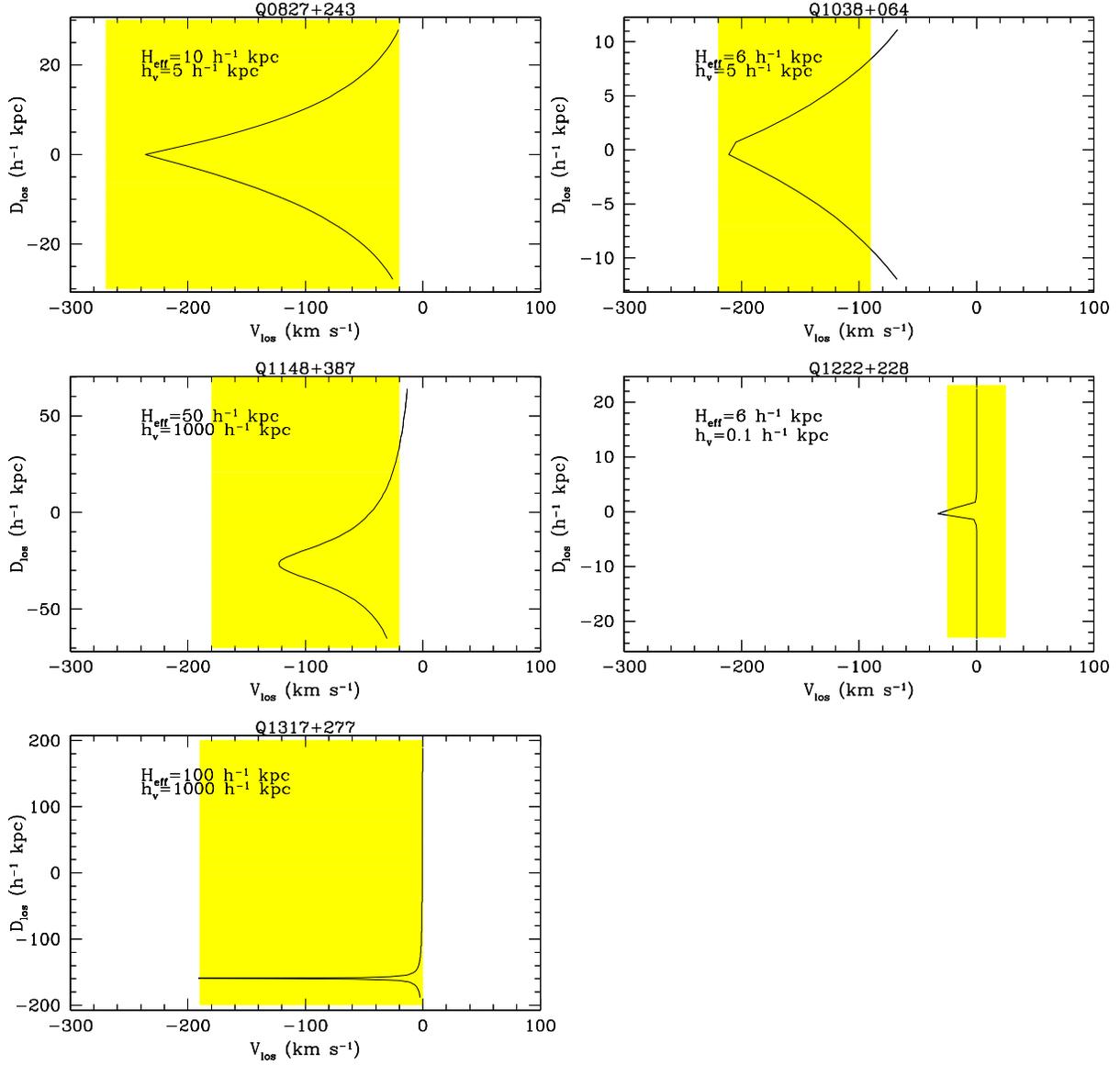}
\caption{Expected velocity as a function of position along the LOS for a)G1 0827+243 b) G1 1038+064 c) G1 1148+387 d)G1 1222+228
and e) G2 1317+276. 
The model parameters used to generate the velocity curves are summarized in table 5. The distance along
the line of sight ${\rm D_{los}}$ is set such that 
the intersection with an extrapolation of the disk midplane occurs at zero. The range of velocities
observed for the Mg~II absorption in each case is indicated by the shaded region.} 
\end{figure}
\end{document}